\documentclass[format=sigconf]{acmart}
\usepackage[labelfont=bf]{caption}

\setcopyright{none}
\renewcommand\footnotetextcopyrightpermission[1]{} % removes footnote with conference information in first column
\acmDOI{}
\acmISBN{}
\acmYear{}

\begin{document}

\title[An Experimental Evaluation of a De-biasing Intervention]{An Experimental Evaluation of a De-biasing Intervention for Professional Software Developers}

\author{Martin Shepperd}
\affiliation{%
  \institution{Brunel University London}
  \city{London}
  \postcode{UB8 3PH}
  \country{UK}
}
\email{martin.shepperd@brunel.ac.uk}

\author{Carolyn Mair}
\affiliation{%
  \institution{Fashion.Psychology}
  \city{London}
  \country{UK}
}
\email{carolyn.mair@gmail.com}

\author{Magne J{\o}rgensen}
\affiliation{%
  \institution{Simula Research Laboratory}
  \city{Oslo}
  \country{Norway}
}
\email{magnej@simula.no}

\begin{abstract}
\emph{\textbf{Context}}: The role of expert judgement is essential in our quest to improve software project planning and execution. However, its accuracy is dependent on many factors, not least the avoidance of judgement biases, such as the anchoring bias, arising from being influenced by initial information, even when it's misleading or irrelevant.  This strong effect is widely documented.\newline
\emph{\textbf{Objective}}:  We aimed to replicate this anchoring bias using professionals and, novel in a software engineering context,  explore de-biasing interventions through increasing knowledge and awareness of judgement biases. \newline
\emph{\textbf{Method}}: We ran two series of experiments in company settings with a total of 410 software developers. Some developers took part in a workshop to heighten their awareness of a range of cognitive biases, including anchoring. Later, the anchoring bias was induced by presenting low or high productivity values, followed by the participants' estimates of their own project productivity.  Our hypothesis was that the workshop would lead to reduced bias, i.e., work as a de-biasing intervention. \newline
\emph{\textbf{Results}}: The anchors had a large effect (robust Cohen's $d=1.19$) in influencing estimates. This was substantially reduced in those participants who attended the workshop (robust Cohen's $d=0.72$). The reduced bias related mainly to the high anchor. The de-biasing intervention also led to a threefold reduction in estimate variance.\newline
\emph{\textbf{Conclusion}}: The impact of anchors upon judgement was substantial.  Learning about judgement biases does appear capable of mitigating, although not removing, the anchoring bias. The positive effect of de-biasing through learning about biases suggests that it has value.
\end{abstract}


\maketitle

\section{Introduction}
\noindent
Effective management of software projects demands, amongst other things, accurate resource predictions.  For this reason cost or effort modelling has been a major topic of research over many years \cite{Jorg07,Malh15}. However, the preponderance of this research has focused on the development and evaluation of formal predictive models.  In contrast, the role of human experts --- who engage in this process, who make choices about model inputs and outputs --- has been somewhat neglected \cite{Jorg04review,Jorg07forecasting}. 

Human judgement and decision-making has been studied for decades by  cognitive psychologists, e.g.,  the well known work of  Kahneman et  al.~\cite{Tver74,Kahn82}.  An important  finding  is  that  humans  typically  use heuristics, i.e., simple mental strategies which, although sufficient in most circumstances, may lead to poor judgements and decisions in others and software engineering is not exempt from this. 

When the use of heuristics leads to a deviation from a rational norm, such as when the heuristic does not fit the context or is based on misleading or irrelevant input, it leads to errors we call judgement and decision biases.  Heuristics, and consequently the judgement and decision biases, are frequently unconscious. This means that the users of heuristics typically will not be able to explain properly how a judgement and or decision was made, why a poor judgement or decision was made or know how to improve the judgement and decision process.  

Many judgement and decision biases have been identified, however our study focuses on the impact of the anchoring bias.  This bias is thoroughly documented as widespread and leading to significant distortions of judgement \cite{Klay93,Bueh10}.

A judgement based on the anchoring heuristic, e.g., when estimating effort or productivity, may frequently be useful.  Imagine a situation where a technically competent project leader indicates that she believes that a software development task should take about 10 work-hours. You are then asked about giving your judgement about the effort you would need for that task. Given that the project leader is competent, it saves you time and mental effort to base your thinking process on that 10 work-hours as a good starting point, or to compare the current task with other tasks with size of about 10 work-hours to find out whether this is larger or smaller. It may even improve the accuracy of the effort estimate.  But what if the number used by your anchoring heuristics is totally irrelevant, such as the number of hours spent on your previous task, or misleading, such as a very low number of work-hours a technically incompetent client wants you to use?  Several studies suggest that software professionals, like everyone else, are affected by presented numbers, even when they are irrelevant or misleading \cite{Jorg11,Halk12}.  This happens even when professionals are explicitly requested to ignore them \cite{Muss01,Jorg04}.

While there are hundreds of studies on the presence of human biases in judgement and decision making, including many on the anchoring bias, there has not been much research on the impact of increased awareness of cognitive biases on the reduction of such biases (i.e., de-biasing). To investigate this topic, we conducted an experiment where the intervention was a workshop to increase participant awareness of cognitive biases and then compared these results with those of participants from a previously published experiment completing the same task who had not attended the workshop \cite{Jorg12}.  

Another limitation of previous research is that most evidence for the existence of  the  anchoring  bias  comes from student samples and the use of tasks where students have little previous experience.  In contrast, the sample in our study comprised professional software developers who were asked to estimate their own productivity on a task they had previously completed. This, we believe, makes the task more familiar for the subject and increases the relevance of the results to real-world tasks.

The remainder of the paper is organised as follows. First we present related work and supporting evidence for cognitive biases and how this might impact judgement and decision making. Next we describe the two-factor experimental design (low and high anchor, de-biasing and no intervention) experiment. We present the results of our robust statistical analysis, initially from 118 professional participants and then pooled with participants from a set of previous experiments. We conclude by discussing the implications of these results for improving professional judgements and outline some areas for further investigation.

\section{Related Work}
\noindent
The anchoring bias is one of the strongest, easiest to create, robust, long-lasting and studied of the human biases \cite{Furn11}. The most famous study of the anchoring bias involved a rigged wheel of fortune and the question: What percentage of the members of the UN are African countries?  First, the research participants span the wheel, which stopped at 10 or 65 depending on how the wheel was rigged, and were asked whether they thought the percentage African countries in the UN was more than or less than the number on the wheel.  Following that question, the participants were asked to predict the proportion of African countries in the UN. The difference in answers between the two groups was large. Those in the first group (wheel stopping at 10) gave a median prediction of 25\% African countries in the UN, while those in the second group (wheel stopping at 65) gave a median prediction of 45\% \cite{Tver74}. It is hard to imagine that the participants would think that a number on a wheel of fortune, which they believed gave a random number between 0 and 100, revealed any information about the actual proportion of African countries in the UN.  Nevertheless, they were strongly affected by the number presented to them. Numerous subsequent studies, following similar anchoring inducing procedures, have shown similar effects. Even completely irrelevant anchors, such as digits from social security numbers or phone numbers, have been demonstrated to strongly bias people's judgements \cite{Arie03}

The anchoring bias is clearly relevant outside artificial experimental settings.  Software professionals' time predictions were for example strongly affected by knowledge about what a customer had communicated as her expectation of time usage, in spite of being informed that the customer had no competence in predicting the time usage \cite{Jorg04}.  When asking these professionals whether they thought they had been affected by the customer's expectations, i.e., by the anchoring information, they either denied it or responded that they were just affected a little. This feeling of not being much affected, when in reality being affected a lot, is part of what makes the anchoring bias potent and hard to avoid.  Even extreme anchors or suggestions, for instance that the length of a whale is 900 metres (unreasonably high anchor) or 0.2 metres (unreasonably low anchor), is effective in influencing people's judgements \cite{Strac97}. Anchoring effects seem to be pretty robust to all kinds of warnings.  The following are instructions from a software development effort estimation study on anchoring: \emph{I admit I have no experience with software projects, but I guess this will take about two months to finish. I may be wrong, of course; we'll wait for your calculations for a better estimate }\cite{Aran05}.  In spite of the warnings, the software developers were strongly affected by the anchoring value of two months.

The cognitive basis of the anchoring bias is disputed and there are at least three different, partly overlapping, explanations: 1) Anchoring as communication (the attitude change theory), i.e., that it is natural for us to give weight to what other people communicate \cite{Wege01}. 2) Anchors as a starting point (the anchoring and adjustment theory), i.e., that the anchor is the starting point and that the adjustment away from the anchor typically is insufficient \cite{Kahn82}. 3) Anchors as an activating experience (the selective accessibility theory), i.e., that the anchor activates experiences and that recently activated experience is more likely to be used in the subsequent judgement process \cite{Muss99}. All explanations have supporting evidence and it is possible that they all contribute to the observed anchoring bias.

De-biasing is applying mitigating interventions to reduce the impact of a bias.  Fischhoff \cite{Fisc81} suggests a fourfold classification scheme:
\begin{itemize}
\item (a) warning about the possibility of bias without specifying its nature. 
\item (b) describing the direction (and possibly extent) of the bias that might typically be observed.
\item (c) providing feedback, preferably at a personal level.
\item (d) offering an extended program of training with feedback, coaching, etc
\end{itemize}

Given the large effect size and importance of the anchor bias, it is not surprising that research has been devoted to study de-biasing strategies, including how to reduce or remove the anchoring bias.  Although several methods for de-biasing have been proposed and tested, researchers have struggled to remove this effect.  Examples of de-biasing strategies with some positive effect, but far from eliminating the bias, are to ``consider the opposite" \cite{Muss00} and introduction of new, more relevant, anchors (known as re-biasing) \cite{Lohr16}.  The study by Lovallo and Sibony \cite{Lova10} reported that the 25\% companies best at avoiding and reducing decision biases, i.e., better at de-biasing, had a 5.3\% advantage over the 25\% worst (i.e., 6.9\% vs 1.6\% typical ROI). This suggests that de-biasing strategies are of substantial real-world importance.

In our paper, we examine the de-biasing effect of increasing the awareness of the anchoring effect among software developers. The evidence in support of this type of de-biasing is mixed.  A positive, although not very large, effect of a training-based increase of bias awareness, including the anchoring bias, was reported in \cite{More15}.  In contrast, no positive effect was found from teaching-based increase of bias awareness by \cite{Oliv17}.  The study reported by Welsh et al.~\cite{Wels07} found a positive effect from increased bias awareness on the overconfidence bias, but none for the anchoring bias.  The general finding seems to be that increased bias awareness typically has moderate to no effect on how much people are biased in their judgements and decisions \cite{Kahn11}.  No prior studies have, as far as we know, reported on the effect of increased anchoring bias awareness in the context of professional software developers. 

\section{Experimental method}

\subsection{Participants}
\noindent
This study is based upon two series of experiments.  The first was conducted by MJ and involved 292 participants from industry with no workshop (de-biasing) intervention. The second series were conducted by CM and MS with MJ involved for the first experiment of the second series. These experiments replicated the initial experimental design (this is documented in \cite{Jorg12} as Estimation Task 1). In addition, the de-biasing intervention of a workshop was introduced prior to the actual experimental task. Table \ref{Tab:Participant} shows the counts of participants by treatment.  The participants were all professional software developers drawn from a total of 15 companies and seven different countries as indicated by Table \ref{Tab:Country}.  They were recruited as volunteers from companies with whom MJ had previously collaborated.  This was supplemented by attendees from effort estimation workshops delivered by MS and CM in the UK and New Zealand.

\begin{table}[htp]
\begin{center}
\begin{tabular}{|l|r|r|r|}
\hline
Workshop? & High Anchor & Low Anchor & Total \\
\hline
N & 142 &150 & 292 \\
Y & 60 & 58 & 118 \\
Total & 202 & 208 & 410 \\
\hline
\end{tabular}
\caption{Participants by Treatment}
\label{Tab:Participant}
\end{center}
\end{table}

\begin{table}[htp]
\begin{center}
\begin{tabular}{|l|r|}
\hline
Country & Count \\
\hline
Nepal & 59 \\
New Zealand & 18\\
Poland & 92 \\
Romania & 48 \\
United Kingdom & 16\\
Ukraine & 114 \\
Vietnam & 63 \\
Total & 410 \\
\hline
\end{tabular}
\caption{Participants by Country}
\label{Tab:Country}
\end{center}
\end{table}

\subsection{Experimental Design}
\noindent
The participants were randomly allocated to either the high anchor or low anchor group.  Each group was then given separate anchor values.  The low anchor was based on the question ``Do you believe your coding productivity was greater than 1 LOC per hour on your last project?".  By contrast, the high anchor was based on the question ``Do you believe your coding productivity was less than 200 LOC per hour on your last project?".  Participants recorded the response, `Yes' or `No'.  They were then all asked to report their estimate of programming productivity in LOC per hour.  The actual estimates are used for this analysis.  

The de-biasing intervention comprised a 2--3 hour workshop on cognitive bias and estimation given immediately prior to the above task. Participants were introduced to the concept of cognitive bias and given examples from the psychology literature demonstrating the influence of bias on decision making. The biases covered included over-optimism and over-confidence \cite{Wein80}, planning fallacy \cite{Bueh94}, peak-end rule \cite{Kahn99}, dual-process theory \cite{Deut55}, blind spot bias \cite{Hans14} and anchoring \cite{Tver74}. The workshop concluded with a discussion on the influence of bias on prediction and estimating.  In terms of Fischhoff's \cite{Fisc81} classification scheme of de-biasing interventions we (b) described the direction and possible extent of the bias and (c) provided some personal feedback via an example task.

\subsection{Data Collection and Cleaning} \label{Sec:Clean}
\noindent
We recorded the following information from each participant summarised in Table \ref{Tab:DataExplan}.

\begin{table}[htp]
\begin{center}
\begin{tabular}{|p{1.5cm}|p{6cm}|}
\hline
Variable  & Explanation \\
\hline
P\_id & Unique participant id \\
Workshop & Y or N depending on the use of a de-biasing intervention \\
Block &  Specific id of the experiment, e.g., there are multiple deliveries for some companies either at different times or locations. \\
Company & The employing company of the software developer - anonymised \\
Country & The country where the software development company is located \\
Anchor & High or low depending on the randomly allocated treatment \\
EstProd & Estimated coding productivity in LOC per hour for the last completed software project. This is the response variable. \\
\hline
\end{tabular}
\end{center}
\caption{Data Collected}
\label{Tab:DataExplan}
\end{table}%

In terms of data cleaning we discarded participants who estimated their productivity as:
\begin{itemize}
\item missing values (5 cases eliminated)
\item zero values as this implied that the participant had not engaged in coding (3 cases eliminated)
\item excessively high values of $ \geqslant 500$ LOC per hour since this implies an implausible level of productivity of almost one LOC per 7 seconds! (4 cases eliminated)
\end{itemize}

A representative sample of five rows of the data are given in Table \ref{Tab:DataExamples}.  The raw data and \textsf{R} scripts are available from https://doi.org/10.6084/m9.figshare.5414200.v3 .

\begin{table*}[htp]
\begin{center}
\begin{tabular}{|r|l|l|l|l|l|r|}
\hline
 P\_id & Workshop & Block & Company & Country & Anchor & Est\_Prod \\
\hline
P130 & N & G & G & Ukraine & high & 100.0 \\
\hline
P334 & Y & I3 & I & Poland & high & 20.0 \\
\hline
P318 & Y & I3 & I & Poland & low & 1.5 \\
\hline
P250 & N & K & K & Vietnam & low & 15.0 \\
\hline
P10 & N & A & A & Romania & high & 80.0 \\
\hline
\end{tabular}
\end{center}
\caption{Example Data Collected}
\label{Tab:DataExamples}
\end{table*}%

\section{Results}

\subsection{Summary statistics}
\noindent
In this section we present the results of our analysis of the experimental data.  First we give the basic descriptive statistics for the response variable Estimated Productivity, then explore the basic anchoring effect and finally our main intervention: the de-biasing effect of the workshop.

\begin{table}[htp]
\begin{center}
\begin{tabular}{|p{0.8cm}|r|p{0.9cm}|r|r|r|p{0.6cm}|p{0.6cm}|}
\hline
Count & Mean & Median & SD & Min & Max & Trim  & Trim \\
& & & & & & mean & SD \\
\hline
410 & 52.7 & 30 & 58.7 & 0.5 & 300 & 37.5 & 51.0\\
\hline
\end{tabular}
\end{center}
\caption{Summary Statistics for Estimated Productivity}
\label{Tab:SummaryStats}
\end{table}

\begin{figure}[htp]
\begin{center}
\includegraphics[width=\linewidth]{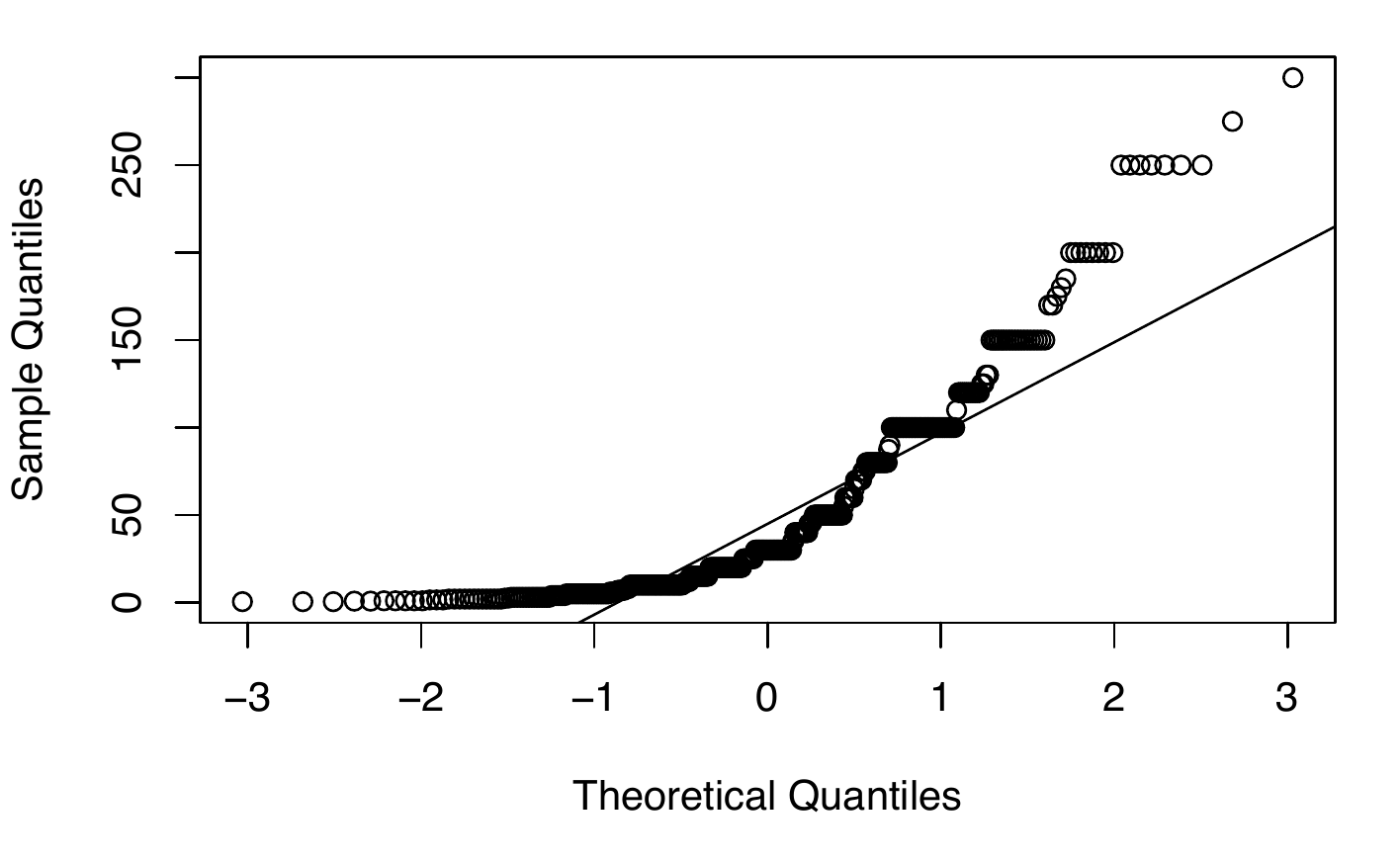}
\caption{Estimated productivity qqplot}
\label{Fig:qqplot}
\end{center}
\end{figure}

Table \ref{Tab:SummaryStats} describes our response variable Estimated Productivity. We see values that range from 0.5 to 300 (after the data cleaning described in Section \ref{Sec:Clean}) with a strong positive skew (evidenced by the mean being greater than the median and the strong deviations particularly of the upper tail in the qqplot (Fig.~\ref{Fig:qqplot}).   For this reason we also compute a 20\% trimmed mean and standard deviation as more robust estimators \cite{Wilc12}.  Both are less than their untrimmed counterparts due to the positive skew (Table \ref{Tab:SummaryStats}).

\begin{figure}[htp]
\begin{center}
\includegraphics[width=\linewidth]{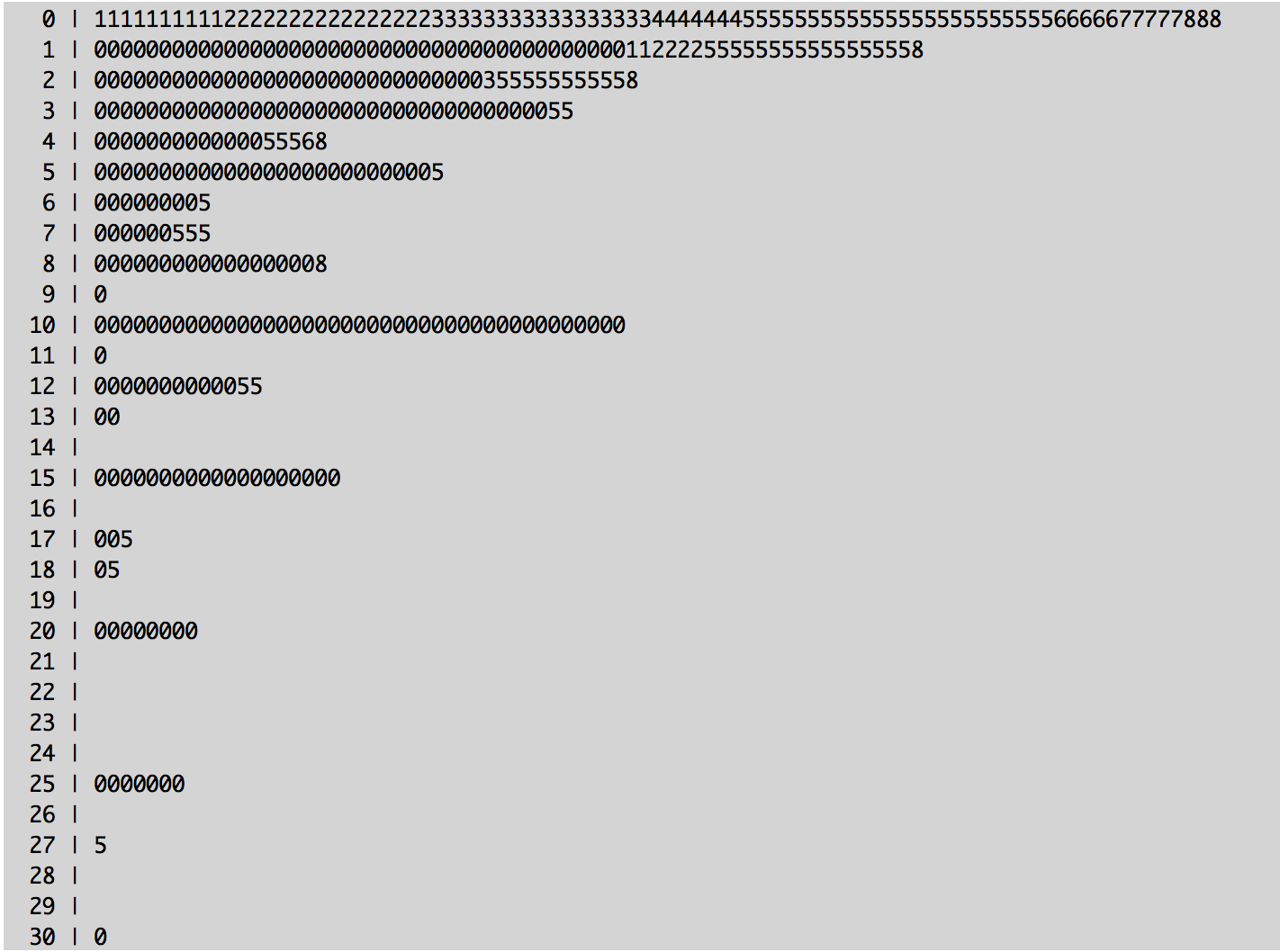}
\caption{Estimated productivity as a stem and leaf plot}
\label{Fig:stem}
\end{center}
\end{figure}

The qqplot also reveals the presence of many ties (horizontal segments of the curve) which correspond to popular round numbers.  For example there are no predictions of 9LOC, but 41 of 10LOC and two of 11LOC.  This is illustrated clearly by the stem and leaf plot where we see zero dominates as a trailing digit, followed by a five (see Fig.~\ref{Fig:stem}).  Perhaps even more remarkable is that not one participant made an estimate ending in a nine.  We conjecture that there is a high degree of uncertainty in the estimates which leads participants to use 5, 10, 20, ... rather than 9 (which would suggest a strong belief in estimation accuracy).  For a discussion of the rounding phenomenon see \cite{Jans01}.

\subsection{The Anchor Effect}
\begin{figure}[htbp]
\begin{center}
\includegraphics[scale=0.55]{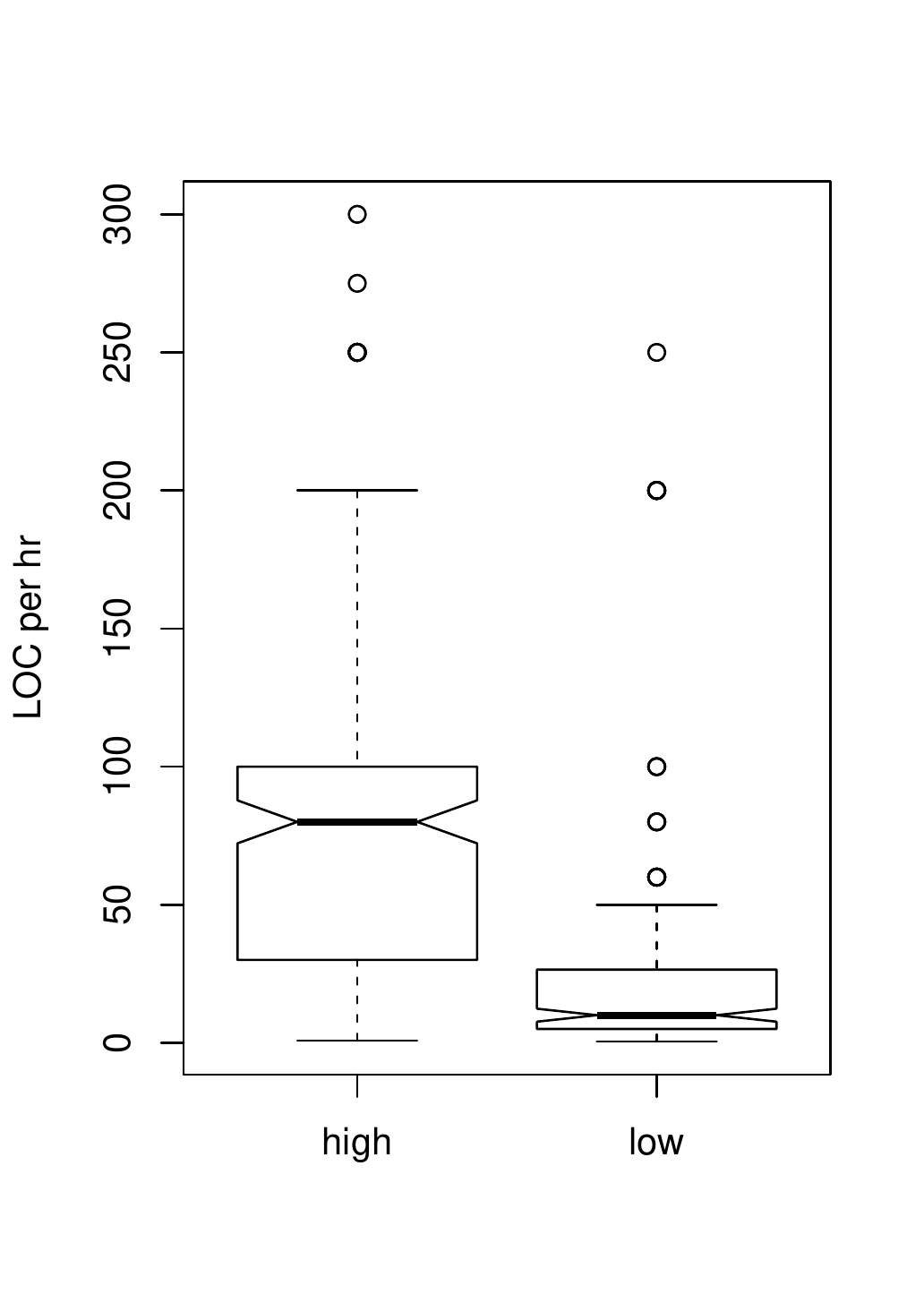}
\caption{Boxplots of Estimated Productivity by Anchor Value}
\label{Fig:AnchorBoxplots}
\end{center}
\end{figure}

\noindent
Recall that we do not know the true productivity levels for each participant.  But given the random allocation of participants to the anchor treatments we do not believe there is any good reason to expect one group to be more productive than the other.   The first thing to observe is the impact of the anchor on all participants, shown graphically in Fig.~\ref{Fig:AnchorBoxplots} as boxplots.  Note the presence of extreme outliers, denoted by individual observations, for both anchor treatments. Note also the substantial difference in medians, shown by the line across each box and the 95\% confidence limits for the medians shown by the notches which do not overlap.

More formally we can compare the two samples using the robust Yuen test with bootstrap to estimate the 95\% confidence interval.  The impact of the anchor is statistically significant, $ p\approx 0$.  The trimmed mean difference is 60.5 and the 95\% confidence interval is (51.6, 69.4).  In terms of effect size, this is either simply the trimmed mean difference of $\sim60$ LOC per hour between a low and high anchor estimate.  Alternatively, if we want to standardise the effect size we can compute a robust version of Cohen's $d$ using a pooled trimmed standard deviation which yields $\sim1.18$, an effect size which is between large and very large (0.8--1.3) \cite{Elli10}. Essentially when software professionals are asked to estimate coding productivity, the percentage difference between the low and high anchor groups was approximately 350\%.

\subsection{The Workshop Effect}
\noindent
So having shown that the anchor effect is very strong in the context of software estimation, we next consider the impact of the de-biasing intervention of the workshop.  But first, we need to address a potential confounder in that the study design is unbalanced; we can see that there are experimental blocks that didn't receive the intervention at all, or vice versa (see Table \ref{Tab:CountryWorkshopCt}). This is potentially problematic as the productivity estimates also differ considerably by country (see Table \ref{Tab:MeanEstProdCountry}).   The UK shows much lower Estimated Productivity and Nepal and Vietnam much higher than the other countries.  Therefore we exclude the UK, Nepal and Vietnam to mitigate this problem.  This leaves 272 participants with 102 receiving the de-biasing intervention.  

\begin{table}[ht]
\centering
\begin{tabular}{|l|r|r|}
  \hline
 Country & N & Y \\ 
  \hline
Nepal &  59 &   0 \\ 
  NZ &   0 &  18 \\ 
  Poland &  47 &  45 \\ 
  Romania &  48 &   0 \\ 
  UK &   0 &  16 \\ 
  Ukraine &  75 &  39 \\ 
  Vietnam &  63 &   0 \\   
\hline
\end{tabular}
\caption{Frequency Count of De-biasing Treatment by Country}
\label{Tab:CountryWorkshopCt}
\end{table}

\begin{table}[htp]
\begin{center}
\begin{tabular}{|l|r|}
\hline
Country & Mean EstProd\\
\hline
Nepal & 75.39 \\	
NZ & 33.4  \\		
Poland & 41.0 \\
Romania & 51.4 \\
UK & 14.6 \\
Ukraine & 47.0 \\
Vietnam & 75.3	\\
\hline
\end{tabular}
\end{center}
\caption{Mean Estimated Productivity by Country}
\label{Tab:MeanEstProdCountry}
\end{table}

%In order to compare what are typical estimates (with and without the workshop) we use the Yuen test of trimmed mean difference, this is 26 with $ p\approx 0$ with a 95\% confidence interval (16, 36).  This leads us to conclude the effect of the de-biasing workshop is to reduce the gap between the high and low anchor by $\sim26$ LOC per hr and the robust Cohen d is $\sim0.72$ i.e., closer to large than medium.

\begin{figure}[htbp]
\begin{center}
\includegraphics[width=\linewidth]{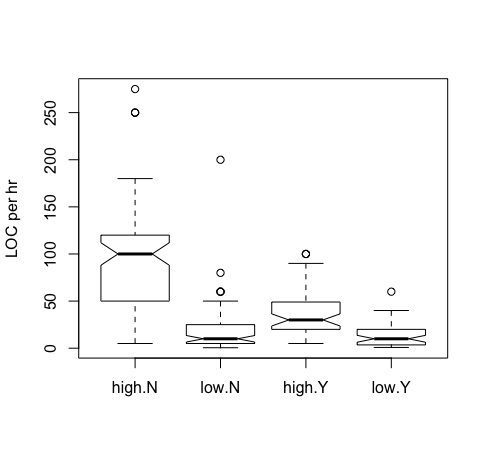}
\caption{Boxplots of Estimated Productivity by De-biasing Intervention\\ \scriptsize{\textbf{Legend:} high.N = high anchor, no de-biasing; low.N = low anchor, no de-biasing; high.Y = high anchor, de-biasing; low.Y = low anchor, de-biasing}}
\label{Fig:AnchorWorkshopBoxplots}
\end{center}
\end{figure}

We compare the estimates visually in Fig.~\ref{Fig:AnchorWorkshopBoxplots} that groups participants both by anchor (low or high) and by de-biasing treatment (Y or N).   It is clear from the boxplots that the median Estimated Productivity for the high anchor without de-biasing (high.N) is substantially greater than the median with de-biasing (high.N).  Recall that the notches indicate the 95\% confidence limits and note that these do not overlap.  As indicated by the size of the whiskers, the spread of estimates also seems greater when there is no de-biasing workshop.   Likewise, we see extreme outliers particularly for the no workshop condition. However, the effect for the low anchor is less obvious.

The median estimate of hourly productivity for the high anchor is reduced from 100 to 30 LOC/hr but for the low anchor the median remains unchanged at 10 LOC/hr. There are three possible reasons for the similarity of the median estimates for those in the low anchor group.  First, the companies, and their software professionals, in the workshop group may have been more productive and as a consequence produced even lower estimates in a no workshop context.  Second, it is harder to influence people to be negative about one's own performance, i.e., that there is less room for de-biasing interventions for the low anchor. Third, the de-biasing intervention may have increased their awareness of the optimism-inducing effect of anchor values, which in this case is the increase in productivity values through a high anchor, but not so much the optimism-reducing effect, corresponding to a low productivity anchor. More studies are needed to analyse and better understand this potentially interesting finding.

\begin{table}[ht]
\centering
\begin{tabular}{|l|c|c|}
  \hline
 Anchor & No workshop & Workshop \\ 
  \hline
high & 92.85 & 37.73 \\  
 & (53.72) &  (27.34)\\  
  low & 19.17 & 13.20 \\  
  & (25.89) &  (12.67)\\ 
   \hline
\end{tabular}
\caption{Mean and Standard Deviations for Estimated Productivity by Anchor and De-biasing Workshop} 
\label{Tab:AnchorWorkshopMeanSD}
\end{table}

We also tabulate comparisons of means and, in parentheses, standard deviations in Table \ref{Tab:AnchorWorkshopMeanSD} and robust analogues based on 20\% trimming in Table \ref{Tab:AnchorWorkshopRobustMeanSD}.  Since trimming tends to remove extreme values we see the general effect is to slightly reduce our estimates of centre and dispersion.

\begin{table}[ht]
\centering
\begin{tabular}{|r|c|c|}
  \hline
 Anchor & No workshop & Workshop \\ 
 \hline
high & 86.44 & 31.42 \\   
 &  (58.51) &  (22.20)\\  
  low & 13.08 & 10.18 \\  
 & (14.05) &  (10.26)\\   
   \hline
\end{tabular}
\caption{20\% Trimmed Mean and Standard Deviations for Estimated Productivity by Anchor and De-biasing Workshop} 
\label{Tab:AnchorWorkshopRobustMeanSD}
\end{table}

Formally we can compare the central tendency and dispersion of the two conditions.  For central tendency we apply the robust Yuen's test and find the trimmed mean difference is 25.6, $ p\approx 0$ and the 95\% confidence interval is (15.5, 35.7).  This strongly suggests that the de-biasing workshop reduces estimates of productivity.  Inasmuch as the higher estimates are influenced upwards by the anchor this is a desirable outcome.

However, we might also expect the spread of estimates to be narrowed if the effect of the anchors are reduced. To compare spread or dispersion we use a simple robust test to compare variance.  We expect the de-biasing to reduce the variance of the estimates since the anchors will have less impact and not stretch out the distribution of estimates.  Robust 20\% trimmed estimates of standard deviation are given in Table \ref{Tab:WorkshopRobustSD} which indicates that the standard deviation is reduced about threefold with the de-biasing workshop intervention.  As a formality we test  that this reduction is significant.  Since we already know the distribution is heavy-tailed, skewed and generally non-Gaussian, we use the Brown-Forsythe median variant of Levene's test of homogeneity of variance \cite{Brow74}.  This gives a Test Statistic of 36.3, $p\approx 0$ meaning it is highly likely the two groups have different variances.
  
\begin{table}[ht]
\centering
\begin{tabular}{|c|c|}
  \hline
 No workshop & Workshop \\ 
 \hline
54.63 & 17.04 \\   
   \hline
\end{tabular}
\caption{20\% Trimmed Standard Deviations for Estimated Productivity by De-biasing Workshop} 
\label{Tab:WorkshopRobustSD}
\end{table}

\begin{figure}[htbp]
\begin{center}
\includegraphics[width=\linewidth]{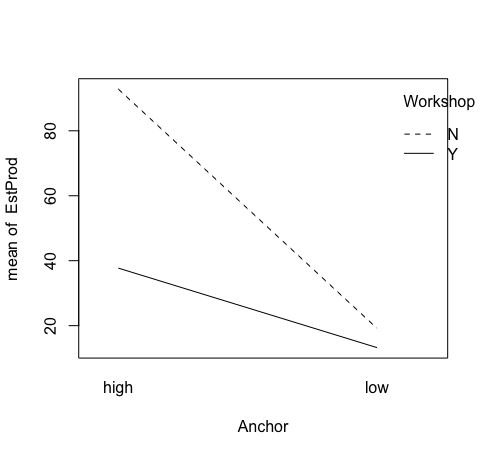}
\caption{Interaction Plot of Anchor and De-biasing Intervention}
\label{Fig:InteractionPlot}
\end{center}
\end{figure}

Considering both factors, the Anchor and the de-biasing Workshop together we use ANOVA, specifically the robust 2-way between-between method of Wilcox \cite{Wilc12,Mair16}.  The results are given in Table \ref{Tab:RobustANOVA} however, we need to sound a note of caution. The variance is strongly heteroscedastic, the data imbalanced and therefore there may be ordering effects, so we only consider gross outcomes.  There is strong evidence that both Anchor and Workshop are associated with estimated productivity, Anchor more so.  It is also clear there is an interaction between Anchor and De-biasing confirmed by the Interaction Plot (Fig.~\ref{Fig:InteractionPlot}).  Essentially the de-biasing intervention only seems to impact the high anchor condition.  This might be because (i) negative values for the estimate are meaningless and (ii) as we suspect the many of the higher values e.g., greater than 100 LOC/hr are somewhat hard to accept.  Therefore it is probable that the high anchor is causing more bias or distortion than the low anchor.

\begin{table}[ht]
\centering
\begin{tabular}{|r|c|c|}
  \hline
 Factor & F & p \\ 
 \hline
Anchor & 192.5 & < 0.001 \\   
Workshop &  72.2 &  < 0.001\\  
Anchor:Workshop & 58.4 & < 0.001 \\  
\hline
\end{tabular}
\caption{Robust 2-way Analysis of Variance} 
\label{Tab:RobustANOVA}
\end{table}

To summarise, we have strong evidence of both the anchor effect and a mitigating effect from the de-biasing workshop.  In terms of effect size, this is either simply the trimmed mean difference of ~26 LOC per hour between an estimate with and without de-biasing.  (This is substantial but less than the Anchor effect).  If we want to standardise we can compute a robust version of Cohen's $d$ using a pooled trimmed standard deviation giving $d \sim 0.72$ which suggests a medium to large effect (0.5 -- 0.8) \cite{Elli10}.  Alternatively the impact of de-biasing can be assessed by considering the reduction in the spread of estimates (since the anchors will have a reducing distorting effect as a consequence of the de-biasing).  We find that the standard deviation of the de-biased estimates is reduced about threefold so again support for the impact of our de-biasing workshops.

\section{Discussion and Conclusions}
\noindent
In this study we have addressed the real world problem of how biases, specifically the anchoring bias, influence software professionals making estimates and then how they might be mitigated.  To do this we have conducted a series of experiments across seven countries with 410 participants.  We believe this study is important because despite the emphasis on formal prediction systems, project cost decisions are ultimately made by humans, and these judgements are infrequent, but of high value.  Therefore they cannot be conceived of as purely technical problems.

Our experiments yield four main findings.
\begin{enumerate}
\item The effect of anchors on software professionals performing estimation tasks, in line with previous studies, such as \cite{Lohr16}, is very strong.
\item The de-biasing workshop significantly reduces --- but does not eliminate --- this bias. 
\item The workshop also substantially reduces the variability in the estimates of professionals approximately threefold
\item The workshop has a greater impact for the high rather than low anchor (although given the meaninglessness of a negative estimate, low estimates could only change in one direction).
\end{enumerate}

However, there are some limitations to this work.  First, we have only considered one type of bias and a relatively simple de-biasing intervention based on a 2-3 hour workshop. There are many other cognitive biases and judgement fallacies, at least some of which could be relevant to software engineering.   Another limitation is that we don't know how long the de-biasing effect will last, but it is quite possible it is only transient.  Therefore follow up work might be useful.

Nevertheless, this study has practical significance.  It shows how professionals can be easily misled into making highly distorted judgements.  This matters in that despite all our tools and automation, software engineering remains a profession that requires judgement and flair.  Fortunately, we show, that it is possible to reduce, although not eliminate, these deleterious effects.   There may well also be considerable scope for refining and improving de-biasing interventions.

\section*{Acknowledgements}
This work was funded by EPSRC Grants EP/I038225/1 and EP/1037881/1.  We are also grateful to the participants of the experiments.

\bibliographystyle{ACM-Reference-Format}
%\balance
\bibliography{SAC2018Refs}


\begin{thebibliography}{37}

%%% the Web address.

\ifx \showCODEN    \undefined \def \showCODEN     #1{\unskip}     \fi
\ifx \showDOI      \undefined \def \showDOI       #1{#1}\fi
\ifx \showISBNx    \undefined \def \showISBNx     #1{\unskip}     \fi
\ifx \showISBNxiii \undefined \def \showISBNxiii  #1{\unskip}     \fi
\ifx \showISSN     \undefined \def \showISSN      #1{\unskip}     \fi
\ifx \showLCCN     \undefined \def \showLCCN      #1{\unskip}     \fi
\ifx \shownote     \undefined \def \shownote      #1{#1}          \fi
\ifx \showarticletitle \undefined \def \showarticletitle #1{#1}   \fi
\ifx \showURL      \undefined \def \showURL       {\relax}        \fi
% invisible to TeX
\providecommand\bibfield[2]{#2}
\providecommand\bibinfo[2]{#2}
\providecommand\natexlab[1]{#1}
\providecommand\showeprint[2][]{arXiv:#2}

\bibitem[\protect\citeauthoryear{Aranda and Easterbrook}{Aranda and
  Easterbrook}{2005}]%
        {Aran05}
\bibfield{author}{\bibinfo{person}{J. Aranda} {and} \bibinfo{person}{S.
  Easterbrook}.} \bibinfo{year}{2005}\natexlab{}.
\newblock \showarticletitle{Anchoring and Adjustment in Software Estimation}.
  In \bibinfo{booktitle}{\emph{ESEC-FSE'05}}. \bibinfo{publisher}{ACM Press},
  \bibinfo{pages}{346--355}.
\newblock


\bibitem[\protect\citeauthoryear{Ariely, Loewenstein, and Prelec}{Ariely
  et~al\mbox{.}}{2003}]%
        {Arie03}
\bibfield{author}{\bibinfo{person}{D. Ariely}, \bibinfo{person}{G.
  Loewenstein}, {and} \bibinfo{person}{D. Prelec}.}
  \bibinfo{year}{2003}\natexlab{}.
\newblock \showarticletitle{``{C}oherent arbitrariness": Stable demand curves
  without stable preferences}.
\newblock \bibinfo{journal}{\emph{The Quarterly Journal of Economics}}
  \bibinfo{volume}{118}, \bibinfo{number}{1} (\bibinfo{year}{2003}),
  \bibinfo{pages}{73--106}.
\newblock


\bibitem[\protect\citeauthoryear{Brown and Forsythe}{Brown and
  Forsythe}{1974}]%
        {Brow74}
\bibfield{author}{\bibinfo{person}{M. Brown} {and} \bibinfo{person}{A.
  Forsythe}.} \bibinfo{year}{1974}\natexlab{}.
\newblock \showarticletitle{Robust tests for the equality of variances}.
\newblock \bibinfo{journal}{\emph{Journal of The American Statistical
  Association}} \bibinfo{volume}{69}, \bibinfo{number}{346}
  (\bibinfo{year}{1974}), \bibinfo{pages}{364--367}.
\newblock


\bibitem[\protect\citeauthoryear{Buehler, Griffin, and Ross}{Buehler
  et~al\mbox{.}}{1994}]%
        {Bueh94}
\bibfield{author}{\bibinfo{person}{R. Buehler}, \bibinfo{person}{D. Griffin},
  {and} \bibinfo{person}{M. Ross}.} \bibinfo{year}{1994}\natexlab{}.
\newblock \showarticletitle{Exploring the `Planning Fallacy': why people
  underestimate their task completion times}.
\newblock \bibinfo{journal}{\emph{Journal of Personality \& Social Psychology}}
  \bibinfo{volume}{67}, \bibinfo{number}{3} (\bibinfo{year}{1994}),
  \bibinfo{pages}{366--381}.
\newblock


\bibitem[\protect\citeauthoryear{Buehler, Peetz, and Griffin}{Buehler
  et~al\mbox{.}}{2010}]%
        {Bueh10}
\bibfield{author}{\bibinfo{person}{R. Buehler}, \bibinfo{person}{J. Peetz},
  {and} \bibinfo{person}{D. Griffin}.} \bibinfo{year}{2010}\natexlab{}.
\newblock \showarticletitle{Finishing on time: When do predictions influence
  completion times?}
\newblock \bibinfo{journal}{\emph{Organizational Behavior and Human Decision
  Processes}} \bibinfo{volume}{111}, \bibinfo{number}{1}
  (\bibinfo{year}{2010}), \bibinfo{pages}{23--32}.
\newblock


\bibitem[\protect\citeauthoryear{Deutsch and Gerard}{Deutsch and
  Gerard}{1955}]%
        {Deut55}
\bibfield{author}{\bibinfo{person}{M. Deutsch} {and} \bibinfo{person}{H.
  Gerard}.} \bibinfo{year}{1955}\natexlab{}.
\newblock \showarticletitle{A study of normative and informational social
  influences upon individual judgment.}
\newblock \bibinfo{journal}{\emph{The Journal of Abnormal and Social
  Psychology}} \bibinfo{volume}{51}, \bibinfo{number}{3}
  (\bibinfo{year}{1955}), \bibinfo{pages}{629--636}.
\newblock


\bibitem[\protect\citeauthoryear{Ellis}{Ellis}{2010}]%
        {Elli10}
\bibfield{author}{\bibinfo{person}{P. Ellis}.} \bibinfo{year}{2010}\natexlab{}.
\newblock \bibinfo{booktitle}{\emph{The Essential Guide to Effect Sizes:
  Statistical Power, Meta-Analysis, and the Interpretation of Research
  Results}}.
\newblock \bibinfo{publisher}{Cambridge University Press}.
\newblock


\bibitem[\protect\citeauthoryear{Fischoff}{Fischoff}{1981}]%
        {Fisc81}
\bibfield{author}{\bibinfo{person}{B. Fischoff}.}
  \bibinfo{year}{1981}\natexlab{}.
\newblock \bibinfo{booktitle}{\emph{Debiasing}}.
\newblock \bibinfo{type}{{T}echnical {R}eport}. \bibinfo{institution}{DECISION
  RESEARCH EUGENE OR}.
\newblock


\bibitem[\protect\citeauthoryear{Furnham and Boo}{Furnham and Boo}{2011}]%
        {Furn11}
\bibfield{author}{\bibinfo{person}{A. Furnham} {and} \bibinfo{person}{H. Boo}.}
  \bibinfo{year}{2011}\natexlab{}.
\newblock \showarticletitle{A literature review of the anchoring effect}.
\newblock \bibinfo{journal}{\emph{The Journal of Socio-Economics}}
  \bibinfo{volume}{40}, \bibinfo{number}{1} (\bibinfo{year}{2011}),
  \bibinfo{pages}{35--42}.
\newblock


\bibitem[\protect\citeauthoryear{Halkjelsvik and J{\o}rgensen}{Halkjelsvik and
  J{\o}rgensen}{2012}]%
        {Halk12}
\bibfield{author}{\bibinfo{person}{T. Halkjelsvik} {and} \bibinfo{person}{M.
  J{\o}rgensen}.} \bibinfo{year}{2012}\natexlab{}.
\newblock \showarticletitle{From origami to software development: A review of
  studies on judgment-based predictions of performance time.}
\newblock \bibinfo{journal}{\emph{Psychological Bulletin}}
  \bibinfo{volume}{138}, \bibinfo{number}{2} (\bibinfo{year}{2012}),
  \bibinfo{pages}{238--271}.
\newblock


\bibitem[\protect\citeauthoryear{Hansen, Gerbasi, Todorov, Kruse, and
  Pronin}{Hansen et~al\mbox{.}}{2014}]%
        {Hans14}
\bibfield{author}{\bibinfo{person}{Katherine Hansen}, \bibinfo{person}{Margaret
  Gerbasi}, \bibinfo{person}{Alexander Todorov}, \bibinfo{person}{Elliott
  Kruse}, {and} \bibinfo{person}{Emily Pronin}.}
  \bibinfo{year}{2014}\natexlab{}.
\newblock \showarticletitle{People Claim Objectivity After Knowingly Using
  Biased Strategies}.
\newblock \bibinfo{journal}{\emph{Personality and Social Psychology Bulletin}}
  \bibinfo{volume}{40}, \bibinfo{number}{6} (\bibinfo{year}{2014}),
  \bibinfo{pages}{691--699}.
\newblock


\bibitem[\protect\citeauthoryear{Jansen and Pollmann}{Jansen and
  Pollmann}{2001}]%
        {Jans01}
\bibfield{author}{\bibinfo{person}{C. Jansen} {and} \bibinfo{person}{M.
  Pollmann}.} \bibinfo{year}{2001}\natexlab{}.
\newblock \showarticletitle{On round numbers: Pragmatic aspects of numerical
  expressions}.
\newblock \bibinfo{journal}{\emph{Journal of Quantitative Linguistics}}
  \bibinfo{volume}{8}, \bibinfo{number}{3} (\bibinfo{year}{2001}),
  \bibinfo{pages}{187--201}.
\newblock


\bibitem[\protect\citeauthoryear{J{\o}rgensen}{J{\o}rgensen}{2004}]%
        {Jorg04review}
\bibfield{author}{\bibinfo{person}{M. J{\o}rgensen}.}
  \bibinfo{year}{2004}\natexlab{}.
\newblock \showarticletitle{A review of studies on expert estimation of
  software development effort}.
\newblock \bibinfo{journal}{\emph{Journal of Systems and Software}}
  \bibinfo{volume}{70}, \bibinfo{number}{1} (\bibinfo{year}{2004}),
  \bibinfo{pages}{37--60}.
\newblock


\bibitem[\protect\citeauthoryear{J{\o}rgensen}{J{\o}rgensen}{2007}]%
        {Jorg07forecasting}
\bibfield{author}{\bibinfo{person}{M. J{\o}rgensen}.}
  \bibinfo{year}{2007}\natexlab{}.
\newblock \showarticletitle{Forecasting of software development work effort:
  Evidence on expert judgement and formal models}.
\newblock \bibinfo{journal}{\emph{International Journal of Forecasting}}
  \bibinfo{volume}{23}, \bibinfo{number}{3} (\bibinfo{year}{2007}),
  \bibinfo{pages}{449--462}.
\newblock


\bibitem[\protect\citeauthoryear{J{\o}rgensen and Grimstad}{J{\o}rgensen and
  Grimstad}{2011}]%
        {Jorg11}
\bibfield{author}{\bibinfo{person}{M. J{\o}rgensen} {and} \bibinfo{person}{S.
  Grimstad}.} \bibinfo{year}{2011}\natexlab{}.
\newblock \showarticletitle{The impact of irrelevant and misleading information
  on software development effort estimates: A randomized controlled field
  experiment}.
\newblock \bibinfo{journal}{\emph{IEEE Transactions on Software Engineering}}
  \bibinfo{volume}{37}, \bibinfo{number}{5} (\bibinfo{year}{2011}),
  \bibinfo{pages}{695--707}.
\newblock


\bibitem[\protect\citeauthoryear{J{\o}rgensen and Grimstad}{J{\o}rgensen and
  Grimstad}{2012}]%
        {Jorg12}
\bibfield{author}{\bibinfo{person}{M. J{\o}rgensen} {and} \bibinfo{person}{S.
  Grimstad}.} \bibinfo{year}{2012}\natexlab{}.
\newblock \showarticletitle{Software Development Estimation Biases: The Role of
  Interdependence}.
\newblock \bibinfo{journal}{\emph{IEEE Transactions on Software Engineering}}
  \bibinfo{volume}{38}, \bibinfo{number}{3} (\bibinfo{year}{2012}),
  \bibinfo{pages}{677--693}.
\newblock


\bibitem[\protect\citeauthoryear{J{\o}rgensen and Shepperd}{J{\o}rgensen and
  Shepperd}{2007}]%
        {Jorg07}
\bibfield{author}{\bibinfo{person}{M. J{\o}rgensen} {and} \bibinfo{person}{M.
  Shepperd}.} \bibinfo{year}{2007}\natexlab{}.
\newblock \showarticletitle{A Systematic Review of Software Development Cost
  Estimation Studies}.
\newblock \bibinfo{journal}{\emph{IEEE Transactions on Software Engineering}}
  \bibinfo{volume}{33}, \bibinfo{number}{1} (\bibinfo{year}{2007}),
  \bibinfo{pages}{33--53}.
\newblock


\bibitem[\protect\citeauthoryear{J{\o}rgensen and Sj{\o}berg}{J{\o}rgensen and
  Sj{\o}berg}{2004}]%
        {Jorg04}
\bibfield{author}{\bibinfo{person}{M. J{\o}rgensen} {and} \bibinfo{person}{D.
  Sj{\o}berg}.} \bibinfo{year}{2004}\natexlab{}.
\newblock \showarticletitle{The impact of customer expectation on software
  development effort estimates}.
\newblock \bibinfo{journal}{\emph{International Journal of Project
  ManagementÊ}} \bibinfo{volume}{22}, \bibinfo{number}{4}
  (\bibinfo{year}{2004}), \bibinfo{pages}{317--325}.
\newblock


\bibitem[\protect\citeauthoryear{Kahneman}{Kahneman}{1999}]%
        {Kahn99}
\bibfield{author}{\bibinfo{person}{D. Kahneman}.}
  \bibinfo{year}{1999}\natexlab{}.
\newblock \bibinfo{booktitle}{\emph{Objective happiness}}.
\newblock \bibinfo{publisher}{Russell Sage Foundation}, \bibinfo{address}{New
  York}, \bibinfo{pages}{3--25}.
\newblock


\bibitem[\protect\citeauthoryear{Kahneman, Lovallo, and Sibony}{Kahneman
  et~al\mbox{.}}{2011}]%
        {Kahn11}
\bibfield{author}{\bibinfo{person}{D. Kahneman}, \bibinfo{person}{D. Lovallo},
  {and} \bibinfo{person}{O. Sibony}.} \bibinfo{year}{2011}\natexlab{}.
\newblock \showarticletitle{Before you make that big decision}.
\newblock \bibinfo{journal}{\emph{Harvard Business Review}}
  \bibinfo{volume}{89}, \bibinfo{number}{6} (\bibinfo{year}{2011}),
  \bibinfo{pages}{50--60}.
\newblock


\bibitem[\protect\citeauthoryear{Kahneman, Slovic, and Tversky}{Kahneman
  et~al\mbox{.}}{1982}]%
        {Kahn82}
\bibfield{author}{\bibinfo{person}{D. Kahneman}, \bibinfo{person}{P. Slovic},
  {and} \bibinfo{person}{A. Tversky}.} \bibinfo{year}{1982}\natexlab{}.
\newblock \bibinfo{booktitle}{\emph{Judgment under uncertainty: Heuristics and
  biases}}.
\newblock \bibinfo{publisher}{Cambridge University Press},
  \bibinfo{address}{Cambridge, UK}.
\newblock


\bibitem[\protect\citeauthoryear{Klayman and Brown}{Klayman and Brown}{1993}]%
        {Klay93}
\bibfield{author}{\bibinfo{person}{J. Klayman} {and} \bibinfo{person}{K.
  Brown}.} \bibinfo{year}{1993}\natexlab{}.
\newblock \showarticletitle{Debias the environment instead of the judge: an
  alternative approach to reducing error in diagnostic (and other) judgment}.
\newblock \bibinfo{journal}{\emph{Cognition}} \bibinfo{volume}{49},
  \bibinfo{number}{1--2} (\bibinfo{year}{1993}), \bibinfo{pages}{97--122}.
\newblock


\bibitem[\protect\citeauthoryear{L{\o}hre and J{\o}rgensen}{L{\o}hre and
  J{\o}rgensen}{2016}]%
        {Lohr16}
\bibfield{author}{\bibinfo{person}{E. L{\o}hre} {and} \bibinfo{person}{M.
  J{\o}rgensen}.} \bibinfo{year}{2016}\natexlab{}.
\newblock \showarticletitle{Numerical anchors and their strong effects on
  software development effort estimates}.
\newblock \bibinfo{journal}{\emph{Journal of Systems and Software}}
  \bibinfo{volume}{116} (\bibinfo{year}{2016}), \bibinfo{pages}{49--56}.
\newblock


\bibitem[\protect\citeauthoryear{Lovallo and Sibony}{Lovallo and
  Sibony}{2010}]%
        {Lova10}
\bibfield{author}{\bibinfo{person}{D. Lovallo} {and} \bibinfo{person}{O.
  Sibony}.} \bibinfo{year}{2010}\natexlab{}.
\newblock \showarticletitle{The case for behavioral strategy}.
\newblock \bibinfo{journal}{\emph{JMcKinsey Quarterly}} \bibinfo{volume}{2010},
  \bibinfo{number}{2} (\bibinfo{year}{2010}), \bibinfo{pages}{30--43}.
\newblock


\bibitem[\protect\citeauthoryear{Mair and Wilcox}{Mair and Wilcox}{2016}]%
        {Mair16}
\bibfield{author}{\bibinfo{person}{P. Mair} {and} \bibinfo{person}{R. Wilcox}.}
  \bibinfo{year}{2016}\natexlab{}.
\newblock \bibinfo{booktitle}{\emph{Robust Statistical Methods in {R}: Using
  the {WRS2} Package}}.
\newblock \bibinfo{type}{{T}echnical {R}eport}. \bibinfo{institution}{Harvard
  University}.
\newblock
\urldef\tempurl%
\url{https://rdrr.io/rforge/WRS2/f/inst/doc/WRS2.pdf}
\showURL{%
\tempurl}


\bibitem[\protect\citeauthoryear{Malhotra}{Malhotra}{2015}]%
        {Malh15}
\bibfield{author}{\bibinfo{person}{R. Malhotra}.}
  \bibinfo{year}{2015}\natexlab{}.
\newblock \showarticletitle{A systematic review of machine learning techniques
  for software fault prediction}.
\newblock \bibinfo{journal}{\emph{Applied Soft Computing}}
  \bibinfo{volume}{27} (\bibinfo{year}{2015}), \bibinfo{pages}{504--518}.
\newblock


\bibitem[\protect\citeauthoryear{Morewedge, Yoon, Scopelliti, Symborski,
  Korris, and Kassam}{Morewedge et~al\mbox{.}}{2015}]%
        {More15}
\bibfield{author}{\bibinfo{person}{C. Morewedge}, \bibinfo{person}{H. Yoon},
  \bibinfo{person}{I. Scopelliti}, \bibinfo{person}{C. Symborski},
  \bibinfo{person}{J. Korris}, {and} \bibinfo{person}{K. Kassam}.}
  \bibinfo{year}{2015}\natexlab{}.
\newblock \showarticletitle{Debiasing decisions: Improved decision making with
  a single training intervention}.
\newblock \bibinfo{journal}{\emph{Policy Insights from the Behavioral and Brain
  Sciences}} \bibinfo{volume}{2}, \bibinfo{number}{1} (\bibinfo{year}{2015}),
  \bibinfo{pages}{129--140}.
\newblock


\bibitem[\protect\citeauthoryear{Mussweiler and Strack}{Mussweiler and
  Strack}{1999}]%
        {Muss99}
\bibfield{author}{\bibinfo{person}{T. Mussweiler} {and} \bibinfo{person}{F.
  Strack}.} \bibinfo{year}{1999}\natexlab{}.
\newblock \showarticletitle{Hypothesis-consistent testing and semantic priming
  in the anchoring paradigm: A selective accessibility model}.
\newblock \bibinfo{journal}{\emph{Journal of Experimental Social Psychology}}
  \bibinfo{volume}{35}, \bibinfo{number}{2} (\bibinfo{year}{1999}),
  \bibinfo{pages}{136--164}.
\newblock


\bibitem[\protect\citeauthoryear{Mussweiler and Strack}{Mussweiler and
  Strack}{2001}]%
        {Muss01}
\bibfield{author}{\bibinfo{person}{T. Mussweiler} {and} \bibinfo{person}{F.
  Strack}.} \bibinfo{year}{2001}\natexlab{}.
\newblock \showarticletitle{The Semantics of Anchoring}.
\newblock \bibinfo{journal}{\emph{Organizational Behavior and Human Decision
  Processes}} \bibinfo{volume}{86}, \bibinfo{number}{2} (\bibinfo{year}{2001}),
  \bibinfo{pages}{234--255}.
\newblock


\bibitem[\protect\citeauthoryear{Mussweiler, Strack, and Pfeiffer}{Mussweiler
  et~al\mbox{.}}{2000}]%
        {Muss00}
\bibfield{author}{\bibinfo{person}{T. Mussweiler}, \bibinfo{person}{F. Strack},
  {and} \bibinfo{person}{T. Pfeiffer}.} \bibinfo{year}{2000}\natexlab{}.
\newblock \showarticletitle{Overcoming the inevitable anchoring effect:
  Considering the opposite compensates for selective accessibility}.
\newblock \bibinfo{journal}{\emph{Personality and Social Psychology Bulletin}}
  \bibinfo{volume}{26}, \bibinfo{number}{9} (\bibinfo{year}{2000}),
  \bibinfo{pages}{1142--1150}.
\newblock


\bibitem[\protect\citeauthoryear{Oliver, Oliver, and Body}{Oliver
  et~al\mbox{.}}{2017}]%
        {Oliv17}
\bibfield{author}{\bibinfo{person}{G. Oliver}, \bibinfo{person}{G. Oliver},
  {and} \bibinfo{person}{R. Body}.} \bibinfo{year}{2017}\natexlab{}.
\newblock \showarticletitle{{BET} 2: Poor evidence on whether teaching
  cognitive debiasing, or cognitive forcing strategies, lead to a reduction in
  errors attributable to cognition in emergency medicine students or doctors}.
\newblock \bibinfo{journal}{\emph{Emergency Medicine Journal}}
  \bibinfo{volume}{34}, \bibinfo{number}{8} (\bibinfo{year}{2017}),
  \bibinfo{pages}{553--554}.
\newblock


\bibitem[\protect\citeauthoryear{Strack and Mussweiler}{Strack and
  Mussweiler}{1997}]%
        {Strac97}
\bibfield{author}{\bibinfo{person}{F. Strack} {and} \bibinfo{person}{T.
  Mussweiler}.} \bibinfo{year}{1997}\natexlab{}.
\newblock \showarticletitle{Explaining the enigmatic anchoring effect:
  Mechanisms of selective accessibility.}
\newblock \bibinfo{journal}{\emph{Journal of Personality and Social
  Psychology}} \bibinfo{volume}{73}, \bibinfo{number}{3}
  (\bibinfo{year}{1997}), \bibinfo{pages}{437--446}.
\newblock


\bibitem[\protect\citeauthoryear{Tversky and Kahneman}{Tversky and
  Kahneman}{1974}]%
        {Tver74}
\bibfield{author}{\bibinfo{person}{A. Tversky} {and} \bibinfo{person}{D.
  Kahneman}.} \bibinfo{year}{1974}\natexlab{}.
\newblock \showarticletitle{Judgment under Uncertainty: Heuristics and Biases}.
\newblock \bibinfo{journal}{\emph{Science}} \bibinfo{volume}{185},
  \bibinfo{number}{4157} (\bibinfo{year}{1974}), \bibinfo{pages}{1124--1131}.
\newblock


\bibitem[\protect\citeauthoryear{Wegener, Petty, Detweiler-Bedell, and
  Jarvis}{Wegener et~al\mbox{.}}{2001}]%
        {Wege01}
\bibfield{author}{\bibinfo{person}{D. Wegener}, \bibinfo{person}{R. Petty},
  \bibinfo{person}{B. Detweiler-Bedell}, {and} \bibinfo{person}{W. Jarvis}.}
  \bibinfo{year}{2001}\natexlab{}.
\newblock \showarticletitle{Implications of attitude change theories for
  numerical anchoring: Anchor plausibility and the limits of anchor
  effectiveness}.
\newblock \bibinfo{journal}{\emph{Journal of Experimental Social Psychology}}
  \bibinfo{volume}{37}, \bibinfo{number}{1} (\bibinfo{year}{2001}),
  \bibinfo{pages}{62--69}.
\newblock


\bibitem[\protect\citeauthoryear{Weinstein}{Weinstein}{1980}]%
        {Wein80}
\bibfield{author}{\bibinfo{person}{N. Weinstein}.}
  \bibinfo{year}{1980}\natexlab{}.
\newblock \showarticletitle{Unrealistic optimism about future life events.}
\newblock \bibinfo{journal}{\emph{Journal of Personality and Social
  Psychology}} \bibinfo{volume}{39}, \bibinfo{number}{5}
  (\bibinfo{year}{1980}), \bibinfo{pages}{806--820}.
\newblock


\bibitem[\protect\citeauthoryear{Welsh, Begg, and Bratvold}{Welsh
  et~al\mbox{.}}{2007}]%
        {Wels07}
\bibfield{author}{\bibinfo{person}{M Welsh}, \bibinfo{person}{S Begg}, {and}
  \bibinfo{person}{R Bratvold}.} \bibinfo{year}{2007}\natexlab{}.
\newblock \showarticletitle{Efficacy of bias awareness in debiasing oil and gas
  judgments}. In \bibinfo{booktitle}{\emph{29th Annual Cognitive Science
  Society}}. Cognitive Science Society, \bibinfo{pages}{1647--1652}.
\newblock


\bibitem[\protect\citeauthoryear{Wilcox}{Wilcox}{2012}]%
        {Wilc12}
\bibfield{author}{\bibinfo{person}{R. Wilcox}.}
  \bibinfo{year}{2012}\natexlab{}.
\newblock \bibinfo{booktitle}{\emph{Introduction to Robust Estimation and
  Hypothesis Testing} (\bibinfo{edition}{3rd} ed.)}.
\newblock \bibinfo{publisher}{Academic Press}.
\newblock


\end{thebibliography}

%-------------------------------------------------------------------------------------------------------------
%\section*{Appendix}

%--------------------------------------------------------------------------------------------------------------

\end{document}